\documentclass{article}
\usepackage[letterpaper,top=2cm,bottom=2cm,left=3cm,right=3cm,marginparwidth=1.75cm]{geometry}
\usepackage{amsmath}
\usepackage{amssymb} 
\usepackage{graphicx}
\usepackage[colorlinks=true, allcolors=blue]{hyperref}
\usepackage{array}
\usepackage{booktabs}
\usepackage{placeins}

\DeclareMathOperator{\Trans}{Transformer}
\DeclareMathOperator{\Token}{Embedding}
\DeclareMathOperator{\Atten}{Attention}
\DeclareMathOperator{\Head}{Head}
\newcommand{\Qbf}{\mathbf{Q}}
\newcommand{\Kbf}{\mathbf{K}}
\newcommand{\Vbf}{\mathbf{V}}
\newcommand{\WQ}{\mathbf{W}_\mathbf{Q}}
\newcommand{\WK}{\mathbf{W}_\mathbf{K}}
\newcommand{\WV}{\mathbf{W}_\mathbf{V}}
\newcommand{\Ebf}{\mathbf{E}}
\newcommand{\Obf}{\mathbf{O}}
\newcommand{\demb}{{d_\text{emb}}}
\newcommand{\dff}{d_{\text{ff}}}

\title{LLM Architecture, Scaling Laws, and Economics:\\
A Quick Summary}
\author{William H. Press\\
Department of Computer Science\\
The University of Texas at Austin}

\begin{document}
\maketitle
\begin{abstract}
The current standard architecture of Large Language Models (LLMs) with QKV self-attention is briefly summarized, including the architecture of a typical Transformer. Scaling laws for compute (flops) and memory (parameters plus data) are given, along with their present (2025) rough cost estimates for the parameters of present LLMs of various scales, including discussion of whether DeepSeek should be viewed as a special case. Nothing here is new, but this material seems not otherwise readily available in summary form.
\end{abstract}

\section{QKV}
Long ago,\cite{salton,deerwester} in the field of information retrieval (IR), some ideas were introduced that later turned out to be seminal: A vector representation of a user's freeform query could be projected into a latent space of possible queries by a matrix $\Qbf$. Correspondingly, a curated set of keywords could be projected, by a matrix $\Kbf$, into the same space, so that $\Qbf \Kbf^T$ then probabilistically mapped queries into keywords. Finally a "values" matrix $\Vbf$ could relate keywords to documents, so that $\Qbf \Kbf^T \Vbf$ (or some variant) mapped queries into retrieved documents.
Later, concepts somewhat analogous also appeared in work on content-based addressing, associative memory, and neural attention \cite{bahdanau,graves,luong}.

This was all background to 
Vaswani et al.'s foundational description of the transformer architecture \cite{vaswani}. Their transformer matrices $\WQ$, $\WK$, $\WV$ may have been inspired by the IR structure, but they lose the specific meanings of ``query", ``keyword", ``value".  Nevertheless they are multiplied together as in the IR usage. It is convenient to view `query", ``keyword", ``value" as mnemonics for the transformer's matrices---but not ruling out a deep structural connection to IR and related subsequent developments. The key difference, however, is that $\WQ$, $\WK$, $\WV$ are all learned from {\em the same} input data, not from any kind of separate ``query", ``keyword", ``value" curations.

Vaswani et al. describe an {\em encoder-decoder} architecture, used for cases that map an input sequence into an output sequence that is not just a continuation of the input. Typical scenarios for this include machine translation, summarization, and speech recognition. LLMs, the most astonishing application of transformers, are {\em decoder-only} architectures: An input stream of words generates a prediction for the next word, and so on, autoregressively. Few would have guessed the power of this model when instantiated at sufficient scale. Whether to call this {\em emergence} is another matter of individual taste. We here discuss just the decoder-only case.

\section{Input/Output Structure}
\label{sec:io}

In large language models (LLMs), words (or words and subwords from byte-pair encoding) are {\em tokenized} into token IDs in some range $1\ldots V$ and then {\em embedded}, each token ID (conceptually with also position information) becoming a real-valued vector of length $\demb$. $V$ is termed the {\em vocabulary size}, while $\demb$ is the {\em embedding-} or {\em model dimension}. (The literature often writes $d_\text{model}$ for our $\demb$.)
\begin{equation}
\text{Embedding } \Ebf : \{1, \dots, V\} \;\to\; \mathbb{R}^{\demb}
\end{equation}
$\Token()$ is a $V \times \demb$ lookup table, $\Ebf$, to be learned along with the rest of the model. Equivalently, $\Ebf$ is a matrix that maps the one-hot encoding of a token ID into its embedding vector by matrix multiplication. The memory required to embed $n$ tokens is $V \times \demb$ (for parameters $\Ebf$) plus $n \times \demb$ activations memory for its output. The operations count is $n V \demb$ if done by one-hot matrix multiplication, or just $n\,\demb$ if by lookup. (Later, we will total up the various memory and operations requirements.)

\begin{figure}
\centering
\includegraphics[width=400pt]{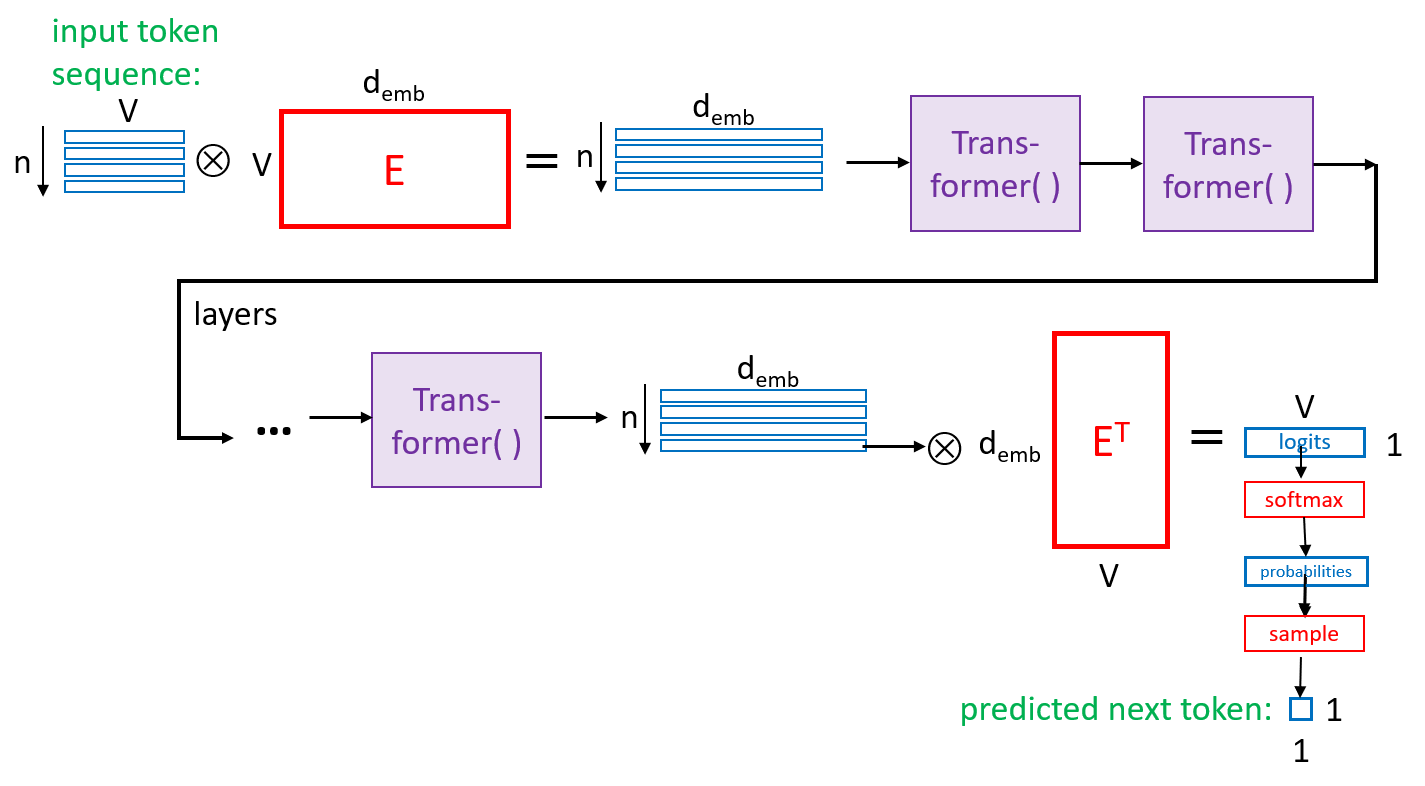}
\caption{\label{fig:LLM}Typical LLM architecture (minus bells and whistles). Input is a sequence of $n$ tokens in a vocabulary $V$, encoded one-hot. Matrix multiplication (here denoted $\otimes$) by an embedding matrix $\Ebf$ produces a corresponding sequence of embeddings. These go through multiple transformer layers. In inference, the last resultant embedding (a single row) is de-embedded into $V$ logits, thence probabilities that may be sampled to produce a predicted next-token. The entire chain is trained simultaneously.}
\end{figure}

The Transformer takes as input a real-valued matrix whose $n$ rows are the embeddings of a sequential stream of $n$ tokens (approximately $n$ words). Rows are arranged along the {\em sequence dimension} whose length $n$ is the {\em context window} length.
The output of the transformer has the same shape as its input. In other words, it transforms:
\begin{equation}
\Trans() : \mathbb{R}^{n \times \demb} \;\to\; \mathbb{R}^{n \times \demb}    
\end{equation}
Transformers can thus be chained into {\em layers}. Schematically, an LLM (simplified) is a tokenization into embeddings, then a chain of $L$ transformer layers, followed by de-embedding of the last-indexed embedding row to yield logit values (after a softmax, probabilities) for each of $V$ possible next tokens. Probabilistic sampling of the last output row produces a single predicted next token (i.e., word). This general scheme is shown in Figure \ref{fig:LLM}.  

The entire chain, $\Ebf$ and all the transformer layers, is trained on a (very) large corpus, by backpropagation and stochastic descent. An important point is that, even though next-token prediction uses only the last ($n$th) output row (see Figure), the training loss function must sum the prediction errors of {\em all} rows. In other words, output token $k$ is trained to predict input token $k+1$ given the previous input tokens $1 \ldots k$. We will understand why below in \S \ref{sec:training}.

\section{The Transformer}
\begin{figure}
\centering
\includegraphics[width=400pt]{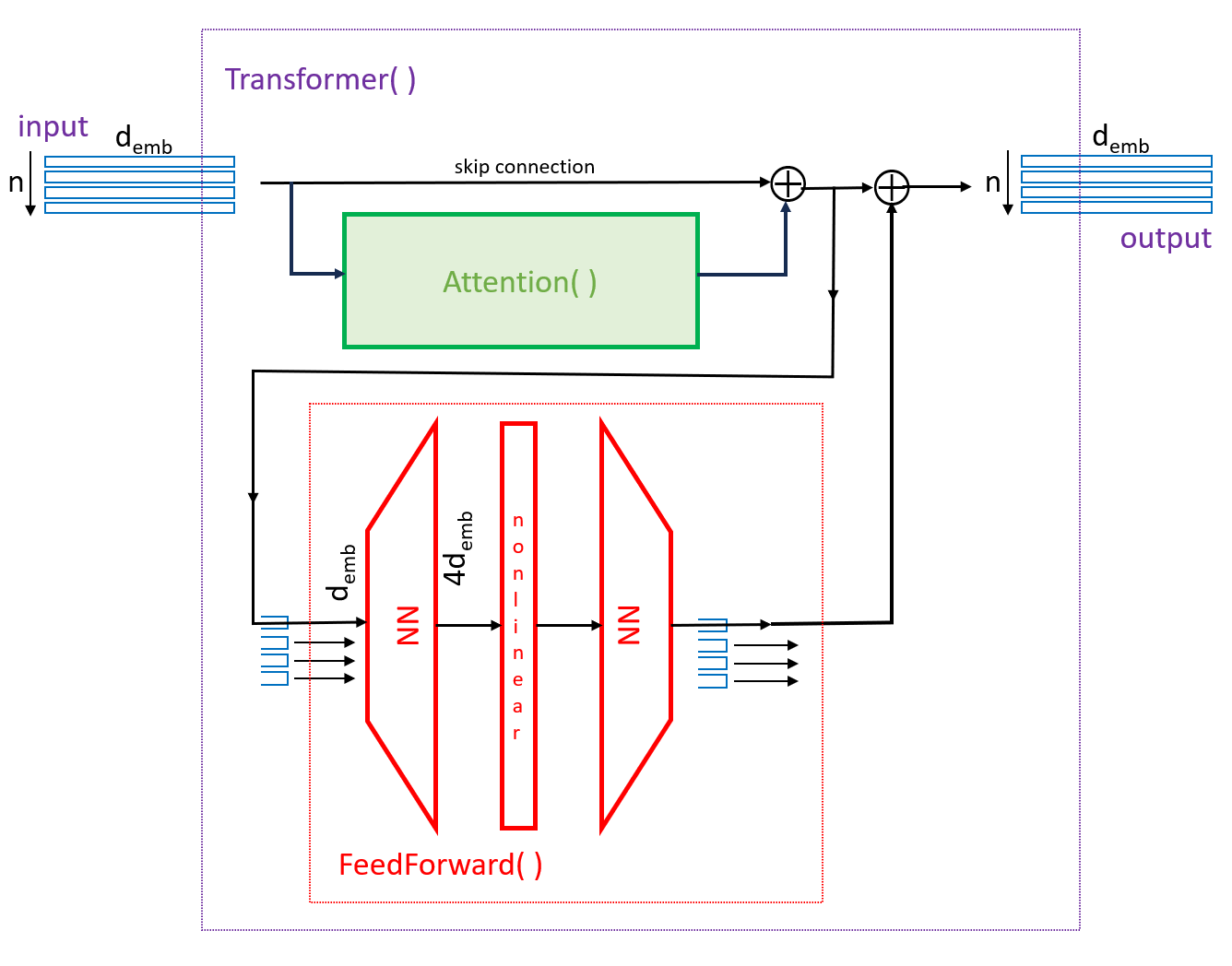}
\caption{\label{fig:Trans}Three components of a typical transformer architecture are a skip conection, an attention function (Figure \ref{fig:attention}), and a feed-forward network. The latter has two fully connected layers separated by a nonlinear activation and is applied separately to each of the $n$ input token embeddings.}
\end{figure}

Now we can dig into the $\Trans( )$ function itself. It combines by additions (see Figure \ref{fig:Trans}) three mappings, all with the identical transformer input and output shapes, $\mathbb{R}^{n \times \demb} \to \mathbb{R}^{n \times \demb}$ : one ``trivial" one, one ``conventional" one, and (the heart of the matter) a very nontrivial one called $\Atten( )$. We consider them in that order, saving the best for last.
\subsection{Skip Connection}
The trivial element of the transformer is a simple pass-through, that is, a {\em skip-connection}. This, or similar, has been a feature in deep neural nets (DNNs) from early days. In forward evaluation, the skip-connection allows a layer (if so trained) to essentially remove itself from the chain of layers. More importantly, in backpropagation, the skip-connection allows gradients to propagate backwards without becoming exponentially small. Nothing about this is specific to transformers.

\subsection{Feed Forward Layer}
The feed-forward layer (FFL) is typically a two-layer conventional NN. Notably, it acts in parallel, with the same parameters, across the sequence dimension, i.e., on each of the $n$ input token embeddings. That is, it doesn't mix across tokens the in sequence. Typically, its first layer is a fully connected expansion from $\demb$ to some $\dff \sim 4\demb$, followed by a nonlinear activation. The second layer is a fully connected contraction back down to $\demb$. The memory footprint is thus $8 \demb^2$. A single evaluation (across all $n$ tokens) has $8 n\,\demb^2 $ flops. (Here and below, we are omitting to mention components with negligible footprint and flops, such as layer-normalization functions, albeit necessary in full implementations.) 

The effect of the feed-forward layer is to create a different token embedding at the output of each $\Trans( )$, compared to what it was at its input: The FFL mixes nonlinearly across the columns of the embedding in the same way for each token, so that all share the new embedding.

\begin{figure}
\centering
\includegraphics[width=440pt]{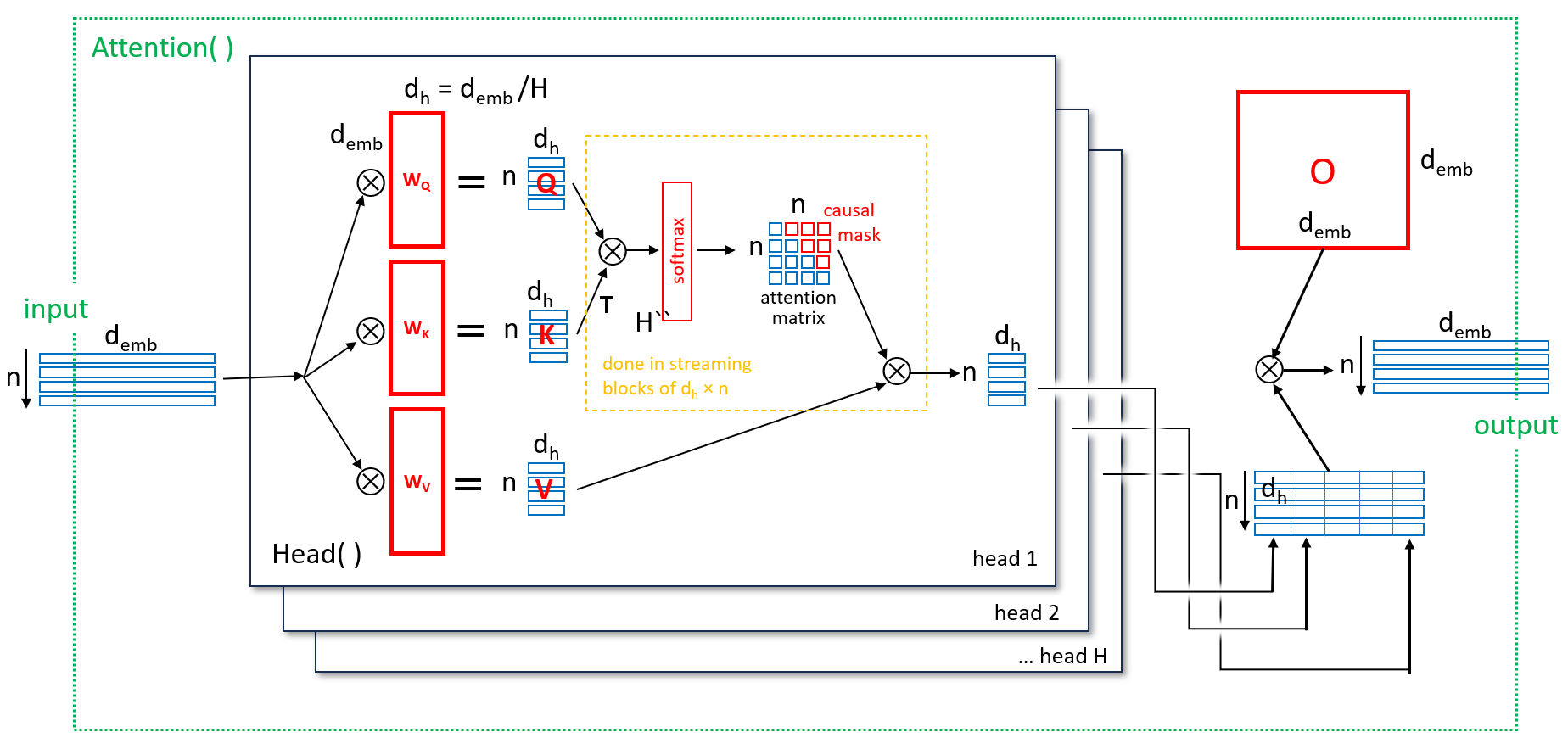}
\caption{\label{fig:attention}The attention function. $H$ attention heads each learn three independent projections from $\demb$ to $d_h$ dimensions. These are combined in each head as $\Qbf\Kbf^T\Vbf$ with a softmax and causal masking (see text). The heads' outputs are concatenated and mapped to output by a fully connected linear layer $\Obf$.}
\end{figure}

\subsection{Attention}
\subsubsection{Multi-headedness}

The $\Atten( )$ function, $\mathbb{R}^{n \times \demb} \to \mathbb{R}^{n \times \demb}$ is composed of many {\em attention heads} (see Figure \ref{fig:attention}). If $H$ is the number of heads, each embodies a mapping into just a subspace of the embedding space,
\begin{equation}
   \Head( ): \mathbb{R}^{n \times \demb} \;\to\; \mathbb{R}^{n \times d_h} 
\end{equation}
where typically $d_h = \demb/H$. In other words, the input and output of a head both span all $n$ tokens (sequence dimension); and the input knows about all $\demb$ embedding dimensions; but the output projects into a smaller attention dimension $d_h$. Even though all heads see the same input, they learn to specialize their outputs usefully. Within a single head, this is called {\em self-attention}; across all the heads it is {\em multi-headed self-attention}.

These multi-head output specializations get put back together in the simplest possible way: concatenated and then multiplied by a $\demb \times \demb$ output matrix $\Obf$ (that is, a fully connected linear layer). See Figure \ref{fig:attention}.

\subsubsection{Q, K, V, finally!}

Within a single attention head, the same $n \times \demb$ input is linearly projected in three separate ways, $\WQ$, $\WK$, and $\WV$  into the smaller $n \times d_h$ matrices $\Qbf,\Kbf,\Vbf$ . 
The symmetry among the three projections is broken under training by the way they are afterward used 
asymmetrically. The outputs $\Qbf$ and $\Kbf$ are contracted (i.e., matrix multiplied) along their $d_h$ dimension producing an $n \times n$ {\em attention matrix}. The attention matrix is then {\em causally masked}, zeroing the implied upper-triangular probabilities so that every input token $k$ ``knows about" only previous tokens.
Notice that this, in each attention head, is the {\em only} all-to-all comparison of tokens in the entire transformer. So, this is where the predicted next token comes to depend on all previous tokens in $H$ (number of heads) separate ways. 

Finally in each head, the computed attention matrix is contracted back to $n \times d_h$ by matrix multiplication with the projection $\Vbf$, ready for concatenation with the other heads (see Figure \ref{fig:attention}).

As for operation count, the  $\WQ$, $\WK$, and $\WV$ matrix multiplies consume $3\, n \,\demb \, d_h$ flops. The product $\Qbf \Kbf^T$ requires $n^2 d_h$, then another $n^2 d_h$ to multiply by $\Vbf$. Since these values are linear in $d_h$, we can get the total over the $H$ heads simply by replacing $d_h$ with $\demb$. That is a key idea in projecting down to multiple heads---the total op count doesn't depend on $H$. Rather, $H$ defines a tradeoff between how many separate ``attentions" there are and how much of the token embedding space each one sees. Multiplication by $\Obf$ consumes another $n\,\demb^2$ flops.

The memory requirement is $4 \,\demb^2$ for $\WQ$, $\WK$, and $\WV$, and $\Obf$ across all the heads, plus, if the above operations are naively done, $4\,n \, \demb + H\, n^2$. That last term is potentially a killer, especially in the case of a large context window, $n \gg \demb$. Luckily (beyond our scope here), the mapping $\Qbf \Kbf^T \Vbf$ can be decomposed into a chunking of submatrix blocks of the attention matrix, requiring memory only on the order of $n \times \demb$, so-called {\em flash attention} \cite{dao}. The full operations count $\propto n^2 \demb$ is still, however, required. Here and below, we are writing $\Qbf \Kbf^T \Vbf$ as shorthand for the more complete $\text{softmax}(\Qbf\Kbf^T/d_h + \mathbf{M})\Vbf$ with $d_h$ a normalizing factor and $\mathbf{M}$ the causal mask in additive logits.)

\begin{table}[h]
\centering
\begin{tabular}{@{}llccc@{}}
\toprule
\shortstack{task \\ { }} & \textbf{ } & \shortstack{flops \\ (mult-adds)} & \shortstack{memory \\ (data)} & \shortstack{memory \\ (parameters)} \\ \midrule
input embedding & token $\to$ embedding & $\approx 0$ & $n \times \demb$ & $V \times \demb$ \\
multi-head attention & projections to $Q,K,V$ & $3\,n\,\demb^2$ & $3\,(n \times \demb)$ & $3\,(\demb\times\demb)$ \\
                     & attention ($\Qbf\Kbf^T\Vbf$) & $(\approx 2)\,n^2 \demb$ & $(\approx 1)\,(n \times \demb)$ &  \\
                     & $\Obf$ projection & $n\,\demb^2$ & $n \times \demb$ & $\demb\times\demb$ \\
feed-forward network & expansion$\times$ 4 & $4\,n\,\demb^2$ & $n \times 4\demb$ & $\demb \times 4\demb$ \\
                     & contraction $\div$ 4 & $4\,n\,\demb^2$ & $n \times \demb$ & $4\demb \times \demb$ \\

layer norms & (not discussed) & $2n\,\demb$ & $2(n \times \demb)$ &  $4\demb$ \\
output projection & $\demb \to V$ logits & $n\,\demb V$ & $n \times V$ & $\demb \times V$ \\
\midrule
\multicolumn{2}{l}{\shortstack{Total (all layers) \\ {} \\ { } \\ { }}}  & \shortstack{$12\,L\,n\,\demb^2$ \\ $+ (\approx 2)L\,n^2\demb$ \\ $+(2L+V)n\,\demb $} &
\shortstack{$(\approx 13)\,n\,\demb$ \\ $+ n\,V$ \\ $\strut$} &
\shortstack{$12\,L\,\demb^2 $ \\ $+ 4L\,\demb$ \\ $+2V\,\demb$ } \\
\bottomrule
\end{tabular}
\caption{Operations, activations memory, and learned parameter counts of $L$ transformer layers in forward evaluation, in terms of context window length $n$, embedding dimension $\demb$, and vocabulary size $V$. Explicit $n \times n$ storage is avoided by chunking (with factors shown as $\approx$). Flop counts score multiply-add (in matrix multiplication) as one flop. Activations memory is reused in each layer, so no factors of $L$ there.}
\label{tab:one}
\end{table}

\section{Flops and Memory Summary}

Table \ref{tab:one} summarizes the order-of-magnitude flops and memory requirements for the forward evaluation of $L$ complete transformer layers, acting on a full context window of $n$ tokens. The total flops are on the order of $2L\,n^2\,\demb + 12L\, n\, \demb^2$. The number of parameters in all transformer layers is about $12L\,\demb^2$, independent of $n$.

\subsection{Autoregressive Forward Evaluation}

In LLMs, the desired response to a user prompt is not a single token or word, but an autoregressive sequence of tokens, each the result of appending the previous one to the context window. It is not necessary to do a full forward evaluation for each token appended.

Rather, modern implementations cache the $\Kbf$ and $\Vbf$ projections of the tokens thus far. When a new token is generated, only its one-row embedding vector must be projected into $\Qbf,\Kbf,\Vbf$, appending only $\Kbf$ and $\Vbf$ to the caches. The cost is $3\,\demb^2$ across all heads. The attention matrix is next augmented by a single row at a cost (for all heads) $n\,\demb$. The product with $\Vbf$ is then computed for that row, with additional cost $n\,\demb$.
The point is that past rows of $\Qbf$ are not used, while past rows of $\Kbf$ and $\Vbf$ are.
Feed-forward and (updated) output projection add another $9\demb^2$. The memory required for caching $\Qbf$ and $\Kbf$ is $2Ln\demb$ since all $L$ layers must be cached.

In summary, the incremental cost of producing one additional output token is about $12\,\demb^2+2\,n\,\demb$ per layer. Which term dominates depends (as before) on whether $n \gg \demb$ or $n \ll \demb$. The activations memory grows in the obvious way as $n$ increases.

\subsection{Training}
\label{sec:training}
A standard rule of thumb \cite{Kaplan, chinchilla1} is that backpropagation of gradients costs $\sim 3 \times$ forward, in flops. Since the whole LLM is trained at once, the above estimates must also be multiplied by $L$, the number of layers.
The so-called {\em Chinchilla rule} \cite{chinchilla1,chinchilla2}, is that about 20 training examples (here, training tokens) per parameter are needed for optimal training of a multi--layer transformer model. This optimal training set size $D^*$ is thus
\begin{equation}
   \text{Optimal Training Set Size (tokens)} \equiv D^* \sim 240\,L\,\demb^2
   + 40\,V\,\demb
\label{eq4}
\end{equation}

Incremental updating via cache is not an option for training, even if partial gradients could be accumulated and averaged. This is because the effect of each new training token-prediction must be backpropagated at a cost
\begin{equation}
    \text{forward-backward flops} \approx (6\,n^2\,\demb + 36\, n\, \demb^2)\,L
    +3\,V\,n\,\demb
\label{eq5}
\end{equation}

A total training cost of equation \eqref{eq4} times \eqref{eq5}, a full forward-backward evaluation for each training token, would be prohibitive. That is why,  
as already mentioned in \S\ref{sec:io}, the training loss function rewards {\em every} transformer output row for predicting its corresponding input---this even though only the last output row will be used in forward inference. This design allows every forward evaluation to produce not just one prediction, but $n$ predictions, whose aggregate loss can be backpropagated in a single pass. Thus dividing by a factor $n$, the total training cost estimate is (omitting for clarity the terms in $V$) about
\begin{equation}
    \text{training flops} \approx 240\,L^2\,\demb^3 (6\,n + 36\,\demb)
\end{equation}

\section{Numerical Examples and Economic Estimates}

Table \ref{tab:two} gives current typical values for $n$, $V$, $\demb$, $H$, and $L$ for four use cases, ranging from a small-scale academic LLM to something like year 2025 state-of-the-art.

\begin{table}[h]
\centering
\begin{tabular}{@{}lccccc@{}}
\toprule
\textbf{Use Case} & \shortstack[ll]{$n$\\ (context)} & \shortstack[ll]{$V$\\ (vocabulary)}& \shortstack[ll]{$\demb$\\(embedding)} & \shortstack[ll]{$H$\\(heads)} & \shortstack[ll]{$L$\\(layers)}  \\ \midrule
\shortstack[l]{A. Small-scale academic model\\$\qquad$(e.g. local LLaMA-7B)} 
  & 2048 
  & 32K
  & 4096 
  & 32 
  & 32 \\
\shortstack[l]{B. Mid-scale enterprise assistant\\$\qquad$(LLaMA-33B, Falcon-40B)} 
  & 4096 
  & 64K
  & 6144 
  & 48 
  & 60 \\
\shortstack[l]{C. Large-scale ca.~2021\\$\qquad$(GPT-3, PaLM-62B class)} 
  & 8192 
  & 80K
  & 12288 
  & 96 
  & 96 \\
\shortstack[l]{D. State-of-the-art (2025)\\$\qquad$(GPT-5, Claude-4 class)} 
  & $\gtrsim 100$K
  & $100$K$-200$K
  & $16$K$-20$K
  & $128-192$ 
  & $120-160$ \\
\bottomrule
\end{tabular}
\caption{Illustrative transformer configurations for different LLM use cases, ranging from small academic-scale to state-of-the-art (2025) frontier systems.}
\label{tab:two}
\end{table}

For each of the use cases in Table \ref{tab:two}, Table \ref{tab:three} gives calculated values for the flop counts for forward and incremental evaluations, the Chinchilla optimal training set size, and the total training cost, expressed in flops, GPU years, and dollars. These last two are at best rough approximations, assuming 300 Tflop/s GPUs with a 2025 capital-dominated cost of \$10,000 per year. These values imply a cost per executed flop of \$$1\times 10^{-18}$. (Full-up operating cost may add another 50\%--100\%, here not included.)

As a sanity check on the calculated table, we note that our calculated range for the training cost for case D (2025 state-of-the-art) is in line with speculation by industry insiders \cite{wsj, epochai}. A separate sanity check is that the incremental flops count per token for case D implies an inference cost of \$1--\$2 per million incremental tokens. This, plus a profit markup, is in rough accord with posted prices for OpenAI's and Anthropic's API calls \cite{openai_api,anthropic_api}.

In Use Case D, the large value for inference cache memory of up to 1.3 T seems borderline implausible, although not impossible. It may be that these largest models adopt a different tradeoff between cache memory and recalculation per output token. Their proprietary details are hard to come by.

\begin{table}[h]
\centering
\begin{tabular}{@{}lcccc@{}}
\toprule
\textbf{Parameter} & \textbf{Use Case A} & \textbf{Use Case B} & \textbf{Use Case C} & \textbf{Use Case D}  \\ \midrule
parameters
  & 6.7 B 
  & 28 B 
  & 175 B
  & $400-800$ B \\
activations memory
  & 180 M 
  & 600 M
  & 1.9 B 
  & $30-90$ B \\
inference cache memory
  & 0.5 B 
  & 3 B
  & 20 B 
  & $0.4-1.3$ T\\
  flops forward
  & $15 \times 10^{12}$ 
  & $125 \times 10^{12}$ 
  & $1.6 \times 10^{15}$ 
  & $80 - 400 \times 10^{15}$ \\
flops incremental
  & $7 \times 10^{9}$ 
  & $30 \times 10^{9}$ 
  & $200 \times 10^{9}$ 
  &  $1 - 2 \times 10^{12}$ \\
optimal training set size
  & $130 \times 10^{9}$ 
  & $540 \times 10^{9}$ 
  & $3.5 \times 10^{12}$ 
  & $8 - 15 \times 10^{12}$ \\
training total flops
  & $2.7 \times 10^{21}$ 
  & $50 \times 10^{21}$ 
  & $2 \times 10^{24}$ 
  & $20 - 100 \times 10^{24}$ \\
training total GPU-years
  & $0.27$ 
  & $5$ 
  & $200$ 
  & $2000 - 10000$ \\
training cost (2025)
  & \$2.7K 
  & \$50K 
  & \$2M 
  & \$20M -- \$100M\\
\bottomrule
\end{tabular}
\caption{Calculated parameters associated with the four use cases in Table \ref{tab:two}. The bottom two lines are rough estimates assuming 300 Tflop/s GPUs at a cost of \$10K per year (capital cost dominating). K, M, B, T denote $10^3$, $10^6$, $10^9$, and $10^{12}$, respectively.}
\label{tab:three}
\end{table}

\FloatBarrier
\section{DeepSeek}
High-Flyer/DeepSeek's 2025 release of its DeepSeek model, especially version V3 \cite{deepseek}, attracted wide popular attention \cite{gibney,roeloffs}. Reportedly, it achieved nearly the performance of state-of-the-art models like Use Case D, above, at a small fraction of those models' training and inference costs. The scaling model in this paper don't directly bear on performance; but it is an interesting check on them to see how well they predict DeepSeek's cost as reported.

DeepSeek incorporates a number of architectural differences from the simple Figure \ref{fig:LLM} above. The main one relevant here is its use of MoE ({\em mixture-of-experts}). In inference, a full forward evaluation is routed through, in effect, values $n=128$K (context length), $\demb = 7168$ (embedding), $H=128$ (heads), and $L=61$ (layers), implying by the formulas above, 38B parameters, approximately in the range of our Use Case B. Reported inference costs are roughly consistent with these parameters. However, DeepSeek's multiple ``experts" are trained with a larger total number of latent parameters, 671B, and most of the training is done with the smaller value $n=32$K. The Chinchilla rule would suggest an optimal training set size $671\text{B}\times 20 \approx 13.4 T$, close to the reported training size of 14.8 T \cite{deepseek}. With these values, our scaling formulas predict a total training cost of $2.9\times 10^{24}$ flops, implying $\approx 2.5$M GPU hours or $\$2.9$M. The reported DeepSeek values are $2.8$M hours and $\$5.6$M. Apart from a higher cost per flop than our assumed value, the agreement is remarkably good. 

We can conclude that DeepSeek's architecture does not somehow evade previous scaling laws. Rather, by utilizing different parameters for training vs.~inference (as enabled by an innovative combination of architectural features), it finds a place to reside within the constraints of those scalings that is especially favorable for price/performance. One might hope for further LLM advances---other than just brute-force scalings of all parameters---to emerge.

\section{Discussion}
Three comments:

1. The cost of GPU compute has been estimated to halve about every 2.0 years \cite{hobbhahn}. One might hope for this evolution for the cost figures in Table \ref{tab:three}.

\noindent 2. As discussed in \S\ref{sec:training}, models are trained on predictions not with the full context-window length $n$, but with, in effect, an approximately uniform distribution of context between 1 and $n$. Since training data is batched (or mini-batched) randomly on each pass, any given token will find itself in a range of context positions, which is good. On the other hand, it seems odd that the training goal (predicting all $n$ positions) differs from the intended use (predicting final token only). Table \ref{tab:three} makes clear that this training-cost savings of a factor $n$ is essential to the economic viability of even the smallest LLM models. For Use Case D, the cost of training every token in full context would be \$100M $\times n \sim \$10^{13}$, about one-third of the U.S.~GDP.

\noindent 3. User prompts to ChatGPT or Claude are said to be prefaced by an invisible ``system prompt" that initializes context and sets guardrails. System prompts are speculated to be anything from a few lines to $\sim 10$ kbytes in length. However, there would seem to be little reason not to use a significant fraction of the stated context-window size $n$. The initial data memory after a forward evaluation on such a prompt is the same for all conversations and can be cached, making subsequent user prompts all incremental in cost (4th row in Table \ref{tab:three}). For conversations with length $\lesssim n$, cost will be about linear in the number of added tokens, as they are in fact priced in the OpenAI and Anthropic APIs.

\subsection*{Acknowledgements}
I had a lot of help from ChatGPT-5, and some from Claude-4, in preparing this summary. I gave each bot one section above at a time and asked for corrections and improvements. This sometimes led to substantial conversations, from which I learned a lot. Apart from finding actual errors, the bots were often oddly and repetitively insistent that I should include small details that I thought better to elide or only mention in passing. One could almost imagine that they regarded their internal details with a sense of ownership! I generally ignored their advice on this minutiae. I take responsibility for all significant errors that remain.

\FloatBarrier
\raggedright
\bibliographystyle{unsrturl}
\bibliography{sample}

\begin{thebibliography}{10}

\bibitem{salton}
Gerard Salton, A.~Wong, and C.~S. Yang.
\newblock {A Vector Space Model for Automatic Indexing}.
\newblock {\em Communications of the ACM}, 18(11):613--620, November 1975.
\newblock \href {https://doi.org/10.1145/361219.361220} {\path{doi:10.1145/361219.361220}}.

\bibitem{deerwester}
Scott Deerwester, Susan~T. Dumais, George~W. Furnas, Thomas~K. Landauer, and Richard Harshman.
\newblock {Indexing by Latent Semantic Analysis}.
\newblock {\em Journal of the American Society for Information Science}, 41(6):391--407, September 1990.
\newblock \href {https://doi.org/10.1002/(SICI)1097-4571(199009)41:6<391::AID-ASI1>3.0.CO;2-9} {\path{doi:10.1002/(SICI)1097-4571(199009)41:6<391::AID-ASI1>3.0.CO;2-9}}.

\bibitem{bahdanau}
Dzmitry Bahdanau, Kyunghyun Cho, and Yoshua Bengio.
\newblock {Neural Machine Translation by Jointly Learning to Align and Translate}.
\newblock In {\em International Conference on Learning Representations (ICLR)}, 2015.
\newblock URL: \url{https://arxiv.org/abs/1409.0473}.

\bibitem{graves}
Alex Graves, Greg Wayne, and Ivo Danihelka.
\newblock {Neural Turing Machines}.
\newblock {\em arXiv preprint arXiv:1410.5401}, 2014.
\newblock URL: \url{https://arxiv.org/abs/1410.5401}.

\bibitem{luong}
Minh-Thang Luong, Hieu Pham, and Christopher~D. Manning.
\newblock {Effective Approaches to Attention-based Neural Machine Translation}.
\newblock In {\em Proceedings of EMNLP 2015}, pages 1412--1421, Lisbon, Portugal, September 2015. Association for Computational Linguistics.
\newblock URL: \url{https://aclanthology.org/D15-1166/}, \href {https://doi.org/10.18653/v1/D15-1166} {\path{doi:10.18653/v1/D15-1166}}.

\bibitem{vaswani}
Ashish Vaswani, Noam Shazeer, Niki Parmar, Jakob Uszkoreit, Llion Jones, Aidan~N. Gomez, {\L}ukasz Kaiser, and Illia Polosukhin.
\newblock {Attention Is All You Need}.
\newblock In {\em Advances in Neural Information Processing Systems 30 (NeurIPS 2017)}, pages 5998--6008, 2017.
\newblock URL: \url{https://papers.neurips.cc/paper/7181-attention-is-all-you-need.pdf}.

\bibitem{dao}
Tri Dao, Daniel~Y. Fu, Stefano Ermon, Atri Rudra, and Christopher R{\'e}.
\newblock {FlashAttention: Fast and Memory-Efficient Exact Attention with IO-Awareness}.
\newblock In {\em Advances in Neural Information Processing Systems 35 (NeurIPS 2022)}, 2022.
\newblock URL: \url{https://proceedings.neurips.cc/paper_files/paper/2022/file/67d57c32e20fd0a7a302cb81d36e40d5-Paper-Conference.pdf}.

\bibitem{Kaplan}
Jared Kaplan, Sam McCandlish, Tom Henighan, Tom~B. Brown, Benjamin Chess, Rewon Child, Scott Gray, Alec Radford, Jeffrey Wu, and Dario Amodei.
\newblock {Scaling Laws for Neural Language Models}.
\newblock {\em arXiv preprint arXiv:2001.08361}, 2020.
\newblock URL: \url{https://arxiv.org/abs/2001.08361}.

\bibitem{chinchilla1}
Jordan Hoffmann, Sebastian Borgeaud, Arthur Mensch, Elena Buchatskaya, Trevor Cai, Eliza Rutherford, Diego de~Las~Casas, Lisa~Anne Hendricks, Johannes Welbl, Aidan Clark, et~al.
\newblock {Training Compute-Optimal Large Language Models}.
\newblock {\em arXiv preprint arXiv:2203.15556}, 2022.
\newblock URL: \url{https://arxiv.org/abs/2203.15556}.

\bibitem{chinchilla2}
Jordan Hoffmann, Sebastian Borgeaud, Arthur Mensch, Elena Buchatskaya, Trevor Cai, Eliza Rutherford, Diego de~Las~Casas, Lisa~Anne Hendricks, Johannes Welbl, Aidan Clark, et~al.
\newblock {Training Compute-Optimal Large Language Models}.
\newblock In {\em Advances in Neural Information Processing Systems 35 (NeurIPS 2022)}, pages 30016--30030, 2022.
\newblock URL: \url{https://proceedings.neurips.cc/paper_files/paper/2022/file/c1e2faff6f588870935f114ebe04a3e5-Paper-Conference.pdf}.

\bibitem{wsj}
Deepa Seetharaman.
\newblock {The Next Great Leap in AI Is Behind Schedule and Crazy Expensive}.
\newblock {\em The Wall Street Journal}, 2024.
\newblock URL: \url{https://www.wsj.com/tech/ai/openai-gpt5-orion-delays-639e7693}.

\bibitem{epochai}
Ben Cottier, Robi Rahman, Loredana Fattorini, Nestor Maslej, Tamay Besiroglu, and David Owen.
\newblock {The Rising Costs of Training Frontier AI Models}.
\newblock {\em arXiv preprint arXiv:2405.21015}, May 2024.
\newblock URL: \url{https://arxiv.org/abs/2405.21015}.

\bibitem{openai_api}
{OpenAI}.
\newblock {API Pricing}.
\newblock \url{https://openai.com/api/pricing/}, 2025.
\newblock Accessed: 2025-09-04.

\bibitem{anthropic_api}
{Anthropic}.
\newblock {Pricing}.
\newblock \url{https://www.anthropic.com/pricing}, 2025.
\newblock Accessed: 2025-09-04.

\bibitem{deepseek}
DeepSeek-AI, Aixin Liu, and et~al.
\newblock Deepseek-v3 technical report, 2025.
\newblock URL: \url{https://arxiv.org/abs/2412.19437}, \href {https://arxiv.org/abs/2412.19437} {\path{arXiv:2412.19437}}.

\bibitem{gibney}
Elizabeth Gibney.
\newblock {China's cheap, open AI model DeepSeek thrills scientists}.
\newblock {\em Nature}, 638(8049):13--14, 2025.
\newblock URL: \url{https://www.nature.com/articles/d41586-025-00229-6}, \href {https://doi.org/10.1038/d41586-025-00229-6} {\path{doi:10.1038/d41586-025-00229-6}}.

\bibitem{roeloffs}
Mary~Whitfill Roeloffs.
\newblock {What Is DeepSeek? New Chinese AI Startup Rivals OpenAI---And Claims It's Far Cheaper}, January 2025.
\newblock URL: \url{https://www.forbes.com/sites/maryroeloffs/2025/01/27/what-is-deepseek-new-chinese-ai-startup-rivals-openai-and-claims-its-far-cheaper/}.

\bibitem{hobbhahn}
Marius Hobbhahn and Tamay Besiroglu.
\newblock {Trends in GPU Price-Performance}, 2022.
\newblock Accessed: 2025-09-04.
\newblock URL: \url{https://epoch.ai/blog/trends-in-gpu-price-performance}.

\end{thebibliography}

\end{document}